\documentclass[10pt]{article}
\usepackage{amsmath,amsfonts,amssymb,amsthm,bbm}
\usepackage{pdfsync}
\usepackage{showlabels}
\usepackage{graphicx}
\usepackage[latin1]{inputenc}
\usepackage{hyperref}
\usepackage[all]{xy}
\usepackage{stmaryrd}
\usepackage[english]{babel}

\newcommand{\Ket}[1]{\left\vert #1\right\rangle}
\newcommand{\Bra}[1]{\left\langle #1\right\vert}

\newcommand{\BraKet}[2]{\left\langle#1\vert #2\right\rangle}

\newcommand{\Mv}[1]{\left\langle #1 \right\rangle}

\begin{document}

\title{\bf Information-theoretic equilibrium and observable thermalisation}

\author{\Large{Fabio Anz\`a and Vlatko Vedral}
\smallskip \\
\small{Atomic and Laser Physics, Clarendon Laboratory, University of Oxford, Parks Road, Oxford, UK} 
}

\author{\Large{Fabio Anz\`a and Vlatko Vedral}
\smallskip \\
\small{Atomic and Laser Physics, Clarendon Laboratory, University of Oxford, Parks Road, Oxford, UK} 
}

\date{\today}

%\affil[2]{Centre for Quantum Technologies, National University of Singapore, 117543 Singapore }

%\affil[3]{Department of Physics, National University of Singapore, 2 Science Drive 3, 117551 Singapore}
%\affil[4]{Center for Quantum Information, Institute for Interdisciplinary Information Sciences, Tsinghua University, 100084 Beijing, China}

%\affil[*]{fabio.anza@physics.ox.ac.uk}

%\affil[+]{these authors contributed equally to this work}

%\keywords{Keyword1, Keyword2, Keyword3}

\maketitle

\begin{abstract}
A crucial point in statistical mechanics is the definition of the notion of thermal equilibrium, which is given as the state that maximises the von Neumann entropy, under the validity of some constraints. Arguing that such a notion can never be experimentally probed, in this paper we propose a new notion of thermal equilibrium, focused on observables rather than on the full state of the quantum system. We characterise such notion of thermal equilibrium for an arbitrary observable via the maximisation of its Shannon entropy and we bring to light the thermal properties that such a principle heralds. The relation with Gibbs ensembles is studied and understood. We apply such a notion of equilibrium to a closed quantum system and show that there is always a class of observables which exhibits thermal equilibrium properties and we give a recipe to explicitly construct them. We bring to light an intimate connection of such a principle with the Eigenstate Thermalisation Hypothesis and provide the first prediction about which observables satisfy it.\\
\end{abstract}

\flushbottom

\thispagestyle{empty}

\section*{Introduction}

To understand under which conditions thermodynamics emerges from the microscopic dynamics is the ultimate goal of statistical mechanics. However, despite the fact that the theory is more than 100 years old, we are still discussing its foundations and its regime of applicability. The ordinary way in which thermal equilibrium properties are obtained, in statistical mechanics, is through a complete characterisation of the thermal form of  the state of the system. One way of deriving such form is by using \textit{Jaynes principle}\cite{Greiner,Uffink,Jaynes1,Jaynes2}, which is the constrained maximisation of von Neumann entropy $S_{\mathrm{vN}} =  \mathrm{Tr} \rho \log \rho$. The main point of Jaynes work was that statistical mechanics can be seen under the light of probabilistic inference, in which one is forced to give prediction about some macroscopic properties of the system, even though the information that we have on the system is not complete. Adopting such point of view, Jaynes showed that the unique state that maximises $S_{\mathrm{vN}}$ (compatibly with the prior information that we have on the system) is our best guess about the state of the system at the equilibrium. The outcomes of such procedure are the so-called \textit{Gibbs ensembles}. \\

In the following we argue that such a notion of thermal equilibrium, \emph{de facto} is not experimentally testable because it gives predictions about all possible observables of the system, even the ones which we are not able to measure. To overcome this issue, we propose a weaker notion of thermal equilibrium, specific for a given observable and therefore experimentally verifiable.\\

 The issue is particularly relevant for the so-called ``Pure states statistical mechanics'' \cite{Pure1,Pure2,Pure3,Pure4,Pure5,Pure6,Pure9,Pure10,Pure11,Pure12,Pure13,Pure14,Pure15,Rei1,Rei2}, which aims to understand how and in which sense thermal equilibrium properties emerge in a closed quantum system, under the assumption that the dynamic is unitary. In the last fifteen years we witnessed a revival of interest in these questions, mainly due to remarkable progresses in the experimental investigation of isolated quantum systems\cite{Exp1,Exp2,Exp3,Exp4,Exp5,Exp6}. Nowadays, the high degree of manipulability and isolation from the environment that we are able to reach makes possible to experimentally investigate such kind of questions and to probe the theoretical predictions.

The starting point of Jaynes' derivation of statistical mechanics is that $S_{\mathrm{vN}}$ is a way of estimating the uncertainty that we have about which pure state the system inhabits. Unfortunately, even though $S_{\mathrm{vN}}$ is undoubtedly an important quantity, we know from quantum information theory that it does not address all kind of ignorance we have about the system. In, it is not the entropy of an observable (though the state is observable); its conceptual meaning is not tied to something that we can measure.

This issue is intimately related with the way we acquire information about a system, i.e. via measurements. The process of measuring an observable $\mathcal{O}$ on a quantum system allows to probe only the diagonal part of the density matrix $\Bra{\lambda_i} \rho \Ket{\lambda_i}$, when this is written in the observable eigenbasis $\left\{\Ket{\lambda_i}\right\}$. For such a reason, from the experimental point of view, it is not possible to assess whether a many-body quantum system is at thermal equilibrium (e.g. Gibbs state $\rho_G$): the number of observables needed to probe all the density matrix elements is too big. In any experimentally reasonable situation we have access only to a few (sometimes just one or two) observables. It is therefore natural to imagine situations in which the outcomes of measurements are compatible with the assumption of thermal equilibrium, while the rest of the density matrix of the system is not. \\

%The idea can be effectively illustrated with a simple example of a free particle in a box. Suppose that, for experimental reasons we only have access to the position observable and we want to know if the particle is at thermal equilibrium. The only information that we can get about the system is the probability distribution of the position: $p(x) \equiv \Bra{x} \rho \Ket{x}$. At the equilibrium (i.e. when the state is the Gibbs state $\rho_G$) such a distribution is constant, however the same answer is true if the particle is in any given eigenstate of the Hamiltonian $\Ket{E}$:

%\begin{align}
%& \Bra{x} \rho_G \Ket{x} = \frac{1}{V} = \left\vert\frac{e^{-i k x}}{\sqrt{V}}\right\vert^2 = |\left\langle x | E \right\rangle |^2
%\end{align}
%where $V$ is the volume of the box. Without other observables it is not possible to probe all the density matrix elements. If our results are, up to experimental uncertainties which are always present, compatible with a constant distribution, we conclude that our observations are compatible with the system being in a thermal state. Adopting a more general perspective, for a many-body quantum system and in any experimentally reasonable situation we will encounter a similar behaviour. We only have access to a small number of observables and this might very well cause the impression that the system is in a thermal state.\\

Despite that, we think that the fact that a distribution is compatible with its thermal counterpart will lead to the emergence of certain thermal properties, concerning the specific observable under scrutiny. Building our intuition on that, we propose a new notion of thermal equilibrium specific for a given observable, experimentally verifiable and which relies on a figure of merit that is not the von Neumann entropy. A good choice for such a figure of merit comes from quantum information theory and it is the Shannon entropy $H_{\mathcal{O}}$ of the eigenvalues probability distribution $\left\{p(\lambda_j) \right\}$ of an observable $\mathcal{O}$. The well-known operational interpretation of $H_{\mathcal{O}}$\cite{Nielsen} matches our needs since it addresses the issue of the knowledge of an observable and it provides a measure for the entropy of its probability distribution.\\

Throughout the paper we will work under the assumption that the Hilbert space of the system has finite dimension and we will refer to the cases in which the Hamiltonian of the system has no local conserved quantities, even though it is possible to address situations where there are several conserved quantities, like integrable quantum systems. Moreover, we will assume that the observable has a pure-point spectrum with the following spectral decomposition $\mathcal{O} = \sum_j \lambda_j \Pi_j$, where $\Pi_j$ is the projector onto the eigenspace defined by the eigenvalue $\lambda_j$. $H_\mathcal{O}$ is the entropy of its eigenvalues probability distribution $p(\lambda_j) \equiv \mathrm{Tr} \left( \rho \Pi_j\right)$

\begin{align}
&H_{\mathcal{O}}[ \hat{\rho} ]  \equiv  -  \sum_{j \, : \, \lambda_j \in \sigma_{\mathcal{O}}} p(\lambda_j) \log p(\lambda_j)
\end{align}

where $\sigma_{\mathcal{O}}$ is the spectrum of $\mathcal{O}$. \\

We propose to define the notion of thermal equilibrium, for an arbitrary but fixed observable $\mathcal{O}$, via a characterisation of the probability distribution of its eigenvalues. We will say that\\

 $\mathcal{O}$ {\it is at  thermal equilibrium when its eigenvalues probability distribution $p(\lambda_j)$ maximises the Shannon entropy $H_{\mathcal{O}}$, under arbitrary perturbations with conserved energy}. We call an observable with such a probability distribution: {\it thermal observable}.\\

It is important to note that such a principle characterises only the probability distribution at equilibrium and it does not uniquely identify an equilibrium state. Given the equilibrium distribution $p(\lambda_j) = p^{\mathrm{eq}}(\lambda_j)$, there will be several quantum states $\rho$ which give the same probability distribution for the eigenvalues $\lambda_j$. In this sense this is weaker a notion of equilibrium, with respect to the ordinary one in which the whole state of the system is required to be an equilibrium state.\\

In the rest of the paper we study the main consequences of the proposed notion of observable-thermal-equilbrium:  its physical meaning and the relation with Gibbs ensembles. These results provide evidences that the proposed notion of equilibrium is able to address the emergence of thermalisation. This is our first result. \\

Furthermore, we study the proposed notion of equilibrium in a closed quantum system and prove that there is a large class of bases of the Hilbert space which always exhibit thermal behaviour and we give an algorithm to explicitly construct them. We dub them \textit{Hamiltonian Unbiased Bases} (HUBs) and, accordingly, we call an observable which is diagonal in one of these bases \textit{Hamiltonian Unbiased Observable} (HUO). The existence and precise characterisation of observables which always thermalise in a closed quantum system is our second result. Furthermore, we investigate the relation between the notion of thermal observable and one of the main paradigms of pure states statistical mechanics: the Eigenstate Thermalisation Hypothesis (ETH)\cite{Berry,Shn,ETH1,ETH2,ETH3,ETH4,ETH5,ETH6,ETH7,ETH8,ETH9,ETH10,ETH11,ETH12}. We find an intimate connection between the concept of HUOs and ETH: the reason why these observables thermalise is precisely because they satisfy the ETH. Hence, with the existence and characterisation of the HUOs we are providing a genuine new prediction about which observables satisfies ETH, for any given Hamiltonian. The existence of this relation between HUOs and ETH is a highly non trivial feature and the fact that we can use it to predict which observables will satisfy ETH is our third result. Indeed, as far as we know, there is no way to predict which observables should satisfy ETH in a generic Hamiltonian system. It is an hypothesis which has to be checked by inspection, case by case.\\

In the conclusive section we summarise the results and discuss their relevance for some open questions.\\ 

\section*{Results}

\subsection*{Information-theoretic equilibrium} 
The request that the equilibrium distribution must be a maximum for $H_{\mathcal{O}}$ can be phrased as a constrained optimisation problem, which can be solved using the Lagrange multiplier technique. The details of the treatment can be found in the Methods section. Two sets of equilibrium equations are obtained and we show in which sense they account for the emergence of thermodynamic behaviour in the observable $\mathcal{O}$. We assume that the only knowledge that we have on the system is the normalisation of the state and the mean value of the energy $\Mv{T}=E_0$, where $T$ is the Hamiltonian of the system. We also make the following assumptions on the Hamiltonian: that it has the following spectral decomposition $T = \sum_{\alpha} E_\alpha T_{\alpha}$, where $T_\alpha \equiv \Ket{E_{\alpha}} \Bra{E_{\alpha}}$ and that its eigenvectors $\left\{ \Ket{E_{\alpha}}\right\}$ provide a full basis of the Hilbert space. We call $\Ket{\psi_n}$ the eigenstates of the density operator, $\rho_n$ are the respective projectors and $q_n$ its eigenvalues. To  describe the state of the system we use the following convenient basis: $\left\{ \Ket{j,s}\right\}$ in which the first index $j$ runs over different eigenvalues $\lambda_j$ of $\mathcal{O}$ and the second index $s$ accounts for the fact that there might be degeneracies and it labels the states within each subspace defined by $\lambda_j$. We also make use of the projectors $\Pi_{js} \equiv \Ket{j,s} \Bra{j,s}$. Furthermore, we are going to need the following quantity $\mathcal{E}_n(j,s)\equiv \Bra{\psi_n} \Pi_{j,s} T \Ket{\psi_n}$, which has the following physical interpretation. $\Pi_{j,s}T$ is the projection of the Hamiltonian onto the eigenstate $\Ket{j,s}$. Hence we understand $\mathcal{E}_n(j,s)$ as the ``average (on $\Ket{\psi_n}$) amount of energy stored in the basis element $\Ket{j,s}$''. Using the overlaps $D^{(n)}_{j,s} \equiv \left\langle j,s|\psi_n \right\rangle$ the expression of the constraints:

\begin{subequations}
\begin{align}
&\mathcal{C}_N \equiv \mathrm{Tr}(\rho) - 1 =  \sum_{n;j,s} q_n \left\vert D_{js}^{(n)}\right\vert^2 -1 \\
&\mathcal{C}_E \equiv \mathrm{Tr}\left(\rho T \right) - E_0 = \sum_{n;j,s} q_n \mathcal{E}_n(j,s) - E_0 \,\,\,.
\end{align}
\end{subequations}

It is important to remark here that the knowledge of $E_0$ is always subject to uncertainty, which we call $\delta E$. In any physically reasonable situation there will be two conceptually different sources of uncertainty: a purely experimental one and an exquisitely quantum one. In this sense, all the states $\rho : \mathrm{Tr} \rho T \in I_0 \equiv \left[ E_0 - \frac{\delta E}{2}, E_0 + \frac{\delta E}{2}\right]$ will be solutions of the constraint equation, up to a certain uncertainty $\delta E$. Even though we do not make any assumption on such a quantity, we note that $\delta E$ is usually assumed to be small on a macroscopic scale but still big enough to host a large number of eigenvalues of the Hamiltonian. \\

Exploiting Lagrange's multipliers technique we obtain four sets of equations. Derivatives with respect to the multipliers 
enforce the validity of the constraints while the derivatives with respect to the field variables give two independent set of equations. 
Using two particular linear combinations of them we obtain the following equilibrium equations (EEs):

\begin{subequations}
\begin{align}
&\overline{\mathcal{E}}_n(j,s) = \mathcal{E}_n(j,s) \label{eq:EE1} \\
&\hspace{-0.5cm} - |D_{js}^{(n)}|^2 \log \left(\sum_{m,s'} q_m  |D_{js'}^{(m)}|^2\right) =  \left( 1-\lambda_N \right)|D_{js}^{(n)}|^2   -  \lambda_E \mathcal{E}_n(j,s) \,\,\, , \label{eq:EE2}
\end{align}
\end{subequations}

where $\lambda_E$ and $\lambda_N$ are the Lagrange multipliers associated to $\mathcal{C}_E$ and $\mathcal{C}_N$, respectively.\\

Our first result is the understanding of the physical consequences of these equations. Together, they account for the emergence of thermal behaviour in $\mathcal{O}$. The first one (\eqref{eq:EE1}) gives the stability
under the flow generated by the Hamiltonian and it implies that the equilibrium distribution $p_{\mathrm{eq}}(\lambda_j)$ has to be be invariant under the unitary dynamics. Indeed, writing the time evolution equation for $p(\lambda_j)$, using the von Neumann equation, we obtain

\begin{align}
&i\hbar \frac{\partial }{\partial t} p(\lambda_j) =\sum_{n,s} q_n \left(\overline{\mathcal{E}}_n(j,s)- \mathcal{E}_n(j,s)\right) \stackrel{eq}{=} 0 \,\,\, ,
\end{align}

where the superscript ``$eq$'' stands for ``at the equilibrium'', i.e. after plugging in the EEs. \\

The second equation (\eqref{eq:EE2}) fixes the functional form of the distribution with respect to the Hamiltonian and to the Lagrange multipliers. It can be shown that
it is responsible for the emergence of a thermodynamical relation between $H_\mathcal{O}$ and the mean value of the energy. Integrating \eqref{eq:EE2} over the whole spectrum of $\mathcal{O}$ we obtain

\begin{align}
&H_{\mathcal{O}}^{eq} = (1-\lambda_N) - \lambda_E \, E_0 .
\end{align}

There is a linear contribution in the mean value of the energy, plus a ``zero-point'' term $H^{(0)}_{\mathcal{O}}=(1-\lambda_N)$. This relation
brings to light the thermodynamical meaning of Shannon entropy $H_{\mathcal{O}}^{eq}$ since such linear dependence on the average 
energy is a distinguishing feature of thermodynamic equilibrium. We note a strong analogy with the properties of von Neumann entropy, 
which acquires thermodynamical relevance once the state of the system is the Gibbs state $\rho_G = \frac{e^{-\beta T}}{\mathcal{Z}}$, where 
$\mathcal{Z} = \mathrm{Tr} e^{-\beta T}$ is the partition function

\begin{center}
\begin{align}
& S_{\mathrm{vN}}(\rho_G) = \log \mathcal{Z} + \beta E_0 \qquad \longleftrightarrow \qquad H_{\mathcal{O}}^{eq} = (1-\lambda_N) - \lambda_E \, E_0\,\,. \label{eq:von}
\end{align}
\end{center}

\subsection*{Relation with statistical mechanics}

We now come to the issue of the relation with the ordinary notion of thermal equilibrium and, therefore, with the Gibbs ensemble. With respect to ours, this is a much more stringent condition because it is a characterisation of the full state of the system. For such a reason, in order to understand if our notion of thermal observable is compatible with the standard characterisation of thermal equilibrium, we need to find the condition under which our criterion gives a complete characterisation of the state of the system. Considering that we are using a maximum-entropy principle, a plausible auxiliary condition is the maximisation of the smallest among all the Shannon entropies. Such a request fully characterises the state of the system because the lowest Shannon entropy of the state is unique and it is the one in which the density matrix of the state is diagonal. Indeed, using the Schur-concavity of the Shannon entropy and the Schur-Horn theorem on eigenvalues of an Hermitian matrix \cite{Nielsen} it is easy to prove that

\begin{align}
& \min_{\mathcal{O} \in \mathcal{A}} H_{\mathcal{O}} (\rho) = S_{\mathrm{vN}} (\rho) \equiv - \mathrm{Tr} \left( \rho \log \rho \right) \,\,\, ,
\end{align}

where $\mathcal{A}$ is the algebra of the observables\cite{Strocchi}. Our minimalist request to maximise the lowest Shannon entropy is translated in the condition of maximising von Neumann entropy. The state which maximises von Neumann's entropy, with the constraint of fixed total energy, is precisely the Gibbs state. It is therefore clear that our proposal constitutes an observable-wise generalisation of the ordinary notion of thermal equilibrium.\\

In the next section we will apply the proposed notion of thermal observable to the so-called ``pure-states statistical mechanics''\cite{Pure1,Pure2,Pure3,Pure4,Pure5,Pure6}, which aims at understanding when, and in which sense, closed quantum system might come to thermal equilibrium. In particular we will investigate the relation with the main paradigm of pure-states statistical mechanics: the ETH.

\subsection*{Closed Quantum Systems - Relation to ETH}

ETH, in its original formulation \cite{ETH1,ETH2,ETH3,ETH4,ETH5,ETH6,ETH7,ETH8}, is an ansatz on the matrix elements of an observable when it is written in the Hamiltonian eigenbasis $\left\{ \Ket{E_{\alpha}}\right\}$:

\begin{align}
&\mathcal{O}_{\alpha \beta}^{\mathrm{ETH}} \approx f_O^{(1)}(\overline{E}) \delta_{\alpha \beta} + e^{-\frac{S(\overline{E})}{2}}f^{(2)}_{\mathcal{O}}(\overline{E},\omega) R_{\alpha \beta} \,\,\, , \label{eq:ETH} 
\end{align}

where $\overline{E} \equiv \frac{E_{\alpha} + E_{\beta}}{2}$, $\omega \equiv E_{\alpha} - E_{\beta}$ while $f_O^{(1)}$ and $f_O^{(2)}$ are smooth functions of their arguments. $S(\overline{E})$ is the thermodynamic entropy at energy $\overline{E}$ defined as $e^{S(\overline{E})} \equiv E \sum_{\alpha} \delta_{\epsilon}(\overline{E}- E_{\alpha})$, where $\delta_{\epsilon}$ is a smeared version of the Dirac delta distribution. $R_{\alpha \beta}$ is a complex random variable with zero mean and unit variance. Furthermore, it is important to remember that ETH by itself does not guarantee thermalisation, we need to impose that the initial state has a small dispersion in the energy eigenbasis\cite{ETH4}. When this is true one says that $\mathcal{O}^{\mathrm{ETH}}$ thermalises in the sense that its dynamically evolving expectation value is close to the microcanonical expectation value defined by the conditions on the average energy

\begin{align}
&\Bra{\psi(t)} \mathcal{O}^{\mathrm{ETH}}\Ket{\psi(t)} \simeq \mathrm{Tr} \left( \mathcal{O}^{\mathrm{ETH}} \rho_{mc} \right)\label{eq:AverTherm}
\end{align}
 where $\rho_{mc} = \rho_{mc}(E_0, \delta E)$ is the microcanonical state defined by the condition on the average value of the energy $\mathrm{Tr} \rho T \in I_0$.

The reason why Eq.(\ref{eq:AverTherm}) becomes true if ETH holds is purely technical, in the following sense. The small dispersion assumption on the initial state is telling us that when we expand the initial state in the energy eigenbasis, we will only have non-zero contributions from a certain energy window:

\begin{align}
&\Ket{\psi_0} = \sum_{\alpha} c_\alpha \Ket{E_\alpha} \simeq \sum_{\alpha : E_{\alpha} \in I_0} c_{\alpha} \Ket{E_{\alpha}} \label{eq:pure} \,\, .
\end{align}

If such dispersion $\delta E$ is small enough, the smooth functions $f_{O}^{(1)},f_{O}^{(2)}$ will not vary too much in such an energy window. Once ETH has been checked by inspection, one can conclude that there is thermalisation of the average value in the sense of Eq.(\ref{eq:AverTherm}). It is important to note that, from the conceptual point of view, the ``small energy-dispersion assumption'' is a key element of the emergence of thermal equilibrium but it has nothing to do with ETH which, by itself, is only the ansatz in Eq.(\ref{eq:ETH}). Nevertheless, this assumption is expected to hold in real experiments because, when working with a many-body quantum system, it is almost impossible to prepare coherent superpositions of states with macroscopically different energies \cite{Rei1,Rei2}. \\

The main intuition behind the emergence of ETH for an observable $\mathcal{O}$ (``Berry conjecture'' \cite{Berry,Shn}) is that for a many-body quantum system the Hamiltonian eigenstates should be so complicated that, when they are written in the eigenbasis of $\mathcal{O}$, their coefficients can be effectively described in term of randomly chosen coefficients and this would lead to the emergence of thermalisation. So far, nothing has been said about which observables we should look at, however it is generally expected that such a statement should be true for local observables. We explicitly show that such mechanism is revealed by the maximisation of Shannon entropy and this allows us to give a prediction about which observables should satisfy ETH. In the next section we will also clarify why we expect local observables to satisfy ETH and we will also provide some general conditions on the Hamiltonian whose validity guarantees that local observables will satisfy ETH.\\

Before we continue we need to define the following short-hand notation: $\sum_{\alpha}' \equiv \sum_{\alpha \in I_0}$, where $\alpha \in I_0$ stands for $\alpha : E_\alpha \in I_0$. 

\subsubsection*{Hamiltonian Unbiased Observables and ETH}

To study the relation with ETH we need to change our perspective. The point of view that we are adopting in the following. One of the key-points behind the ETH is that, in many real cases, the expectation values computed onto the Hamiltonian eigenvectors can be very close to the thermal expectation values. Moreover, when one wants to argue that thermalisation in a closed quantum system arises because of ETH, a main assumption is that the initial pure state of the system $\Ket{\psi_0}$ has a very small energy uncertainty $\Delta E$, with respect to the average energy $E_0$: $\frac{\Delta E}{E_0} \ll 1$. Following these two insights, we take the extreme limit in which $\Delta E = 0$. With such a choice, we are left with an Hamiltonian eigenstate and a constraint equation given by $\mathrm{Tr} \rho H = E_n$. We note that Eq.(\ref{eq:EE1}) is trivially satisfied for an Hamiltonian eigenstate. Hence we assume that $\Ket{\psi} = \Ket{E_\alpha} : E_{\alpha} \in I_0$ and use the solvability of Eq.(\ref{eq:EE2}) as a criterion to look for observables which can be thermal. While this is a very specific choice, we will show that it is enough to unravel some interesting feature regarding the ETH. 

 With such an assumption, the second equilibrium equation becomes

\begin{align}
&\hspace{-0.5cm} - |D_{j}|^2 \log  |D_{j}|^2  =  \left( 1-\lambda_N \right)|D_{j}|^2    -  \lambda_E |D_{j}|^2 E_{\alpha}
\end{align}

Dividing by $|D_{j}|^2$ on each side we obtain a right-hand side which does not depend on the label $j$.
Hence, keeping in mind that $x\log x \to 0$ for $x \to 0$, the most general solution of this equation is a 
constant distribution with support on some subset ($\mathcal{I}_{\alpha} \subset \sigma_{\mathcal{O}}$ 
depending on $E_{\alpha}$) of the spectrum and zero on its complementary:

\begin{equation}
p(\lambda_j)  = \left\{ \begin{array}{ll}
  \frac{1}{d_{\alpha}} & \forall \,\, \lambda_j \in \mathcal{I}_{\alpha} \subset \sigma_{\mathcal{O}} \\
 & \\
 0 & \forall \,\, \lambda_j \in \sigma_{\mathcal{O}} / \mathcal{I}_{\alpha}  \label{eq:const} 
  \end{array} \right. 
\end{equation}

where $d_{\alpha} = \mathrm{dim} \, \mathcal{I}_{\alpha}$ is the number of orthogonal states on which the distribution has non-zero value. The distribution of eigenvalues of $\mathcal{O}$ given by Eq.(\ref{eq:const}) fully agrees with the prediction from the microcanonical ensemble $\rho_{mc}^{(\alpha)}$, defined by the condition in Eq.(\ref{eq:const}). This is true, in particular, for the expectation value

\begin{align}
& \Mv{\mathcal{O}} = \frac{1}{d_{\alpha}} \sum_{j \, : \lambda_j \in \mathcal{I}_{\alpha}} \lambda_j = \mathrm{Tr} \left(\mathcal{O} \, \rho_{mc}^{(\alpha)}\right)
\end{align}

It has to be understood here that Eq.(\ref{eq:const}) is a highly non-trivial condition on $\mathcal{O}$, which is not going to be fulfilled by every observable since it imposes a very specific relation between its eigenstates and the energy eigenvectors.\\

Within the context of quantum information theory, and especially in quantum cryptography, a similar relation has already been studied. Two bases of a $\mathcal{D}-$dimensional Hilbert space ($\mathcal{B}_v \equiv \left\{ \Ket{v_i} \right\}$ and $\mathcal{B}_w \equiv \left\{ \Ket{w_j} \right\}$) are called {\it mutually unbiased bases} (MUB) \cite{MUB1,MUB2,MUB3} when 

\begin{align}
&|\left\langle v_i | w_j \right\rangle|^2 = \frac{1}{\mathcal{D}} \qquad \forall \,\, i,j=1,\ldots, \mathcal{D}
\end{align}

Such a concept is a generalisation, expressed in term of vector bases, of canonically conjugated operators. In other words, each vector of $\mathcal{B}_v$ is completely delocalised in the basis $\mathcal{B}_w$ and viceversa. Here we mention the result about MUBs which matter most for our purposes: given the $2^N$-dimensional Hilbert space of $N$ qubits, there are $2^N+1$ MUBs and we have an algorithm to explicitly find all of them\cite{MUB4}. Therefore, if we are in an energy eigenstate, an observable unbiased with respect to the Hamiltonian basis will always exhibit a microcanonical distribution. This is also true if our state is not exactly an energy eigenstate, it is enough to have a state that has a sufficiently narrowed energy distribution. We now provide a simple argument to prove such a statement. We also note that such condition is closely related to the small dispersion condition briefly discussed before and that it is necessary to guarantee thermalisation, according to the ETH.\\ 

By assumption, the pure state in Eq.(\ref{eq:pure}) has an energy distribution $\left\{p_{\alpha} \equiv |c_{\alpha}|^2 \right\}_{\alpha = 1}^{\mathcal{D}}$ with small dispersion. This implies that its Shannon entropy $H_T$ has a small value, because the profile of the distribution will be peaked around a certain value. For such a reason $H_{T}(\Ket{\psi_0})$ will be much smaller than its maximum value

\begin{align}
&H_{T}(\Ket{\psi_0}) \ll \log \mathcal{D} \label{eq:SmallH}
\end{align}

Moreover, it can be proven that between any pair of MUB, like the Hamiltonian eigenbasis and an HUB, there exists the following entropic uncertainty relation, involving their Shannon entropies:

\begin{align}
&H_T(\Ket{\phi}) + H_{HUB}(\Ket{\phi}) \geq \log \mathcal{D} \qquad \forall \Ket{\phi} \in \mathcal{H} \label{eq:EntUnc}
\end{align}

Putting together Eq.(\ref{eq:SmallH}) and Eq.(\ref{eq:EntUnc}) we obtain that, for all the states with a small energy dispersion, the Shannon entropy of an HUB will always have a value close to the maximum:

\begin{align}
&H_{HUB}(\Ket{\psi_0})  \simeq \log \mathcal{D}
\end{align}

This in turn implies that the distribution of all the HUO will be approximately the same as the one computed on the microcanonical state. 

For such a reason, we now study the properties of a HUO:

\begin{align}
&\mathcal{O} = \sum_{j,s} \lambda_j \Ket{j,s} \Bra{j,s} &&\qquad \BraKet{j,s}{E_{\alpha}} =  \frac{e^{i\theta_{js,\alpha}}}{\sqrt{\mathcal{D}}} \label{eq:Const}
\end{align}

\subsubsection*{HUOs and ETH: diagonal matrix elements}

In order to investigate the relation with ETH we need to study the matrix elements of a HUO in the energy basis:

\begin{align}
&\mathcal{O}_{\alpha \beta}= \frac{1}{\mathcal{D}} \sum_{j,s} \lambda_j e^{i \omega_{js}^{\alpha \beta}}  \,\,\, , \label{eq:off}
\end{align}

with $\omega_{js}^{\alpha \beta} =  (\theta_{js,\beta}- \theta_{js,\alpha})$.
It is straightforward to conclude that its diagonal matrix elements are constant in such a basis and therefore the so-called (Hamiltonian) Eigenstate Expectation values reproduce the 
microcanonical expectation values: 

\begin{align}
&\mathcal{O}_{\alpha \alpha} = \frac{1}{\mathcal{D}} \sum_{j,s} \lambda_j = \frac{1}{\mathcal{D}}\mathrm{Tr} \, \mathcal{O} = \mathrm{Tr} \left(\mathcal{O} \, \rho_{mc}\right) = \Mv{\mathcal{O}}_{mc} \label{eq:EE}
\end{align}

This is the first part of our third result and as it is, it can already be used to explain the emergence of thermalisation in closed quantum system. In \cite{Rei1,Rei2} Reimann proved an important theorem about equilibration of closed quantum systems. He was able to show that under certain experimentally realistic conditions, the mean value of an observable is not much different from its value computed on the time-averaged density matrix, or Diagonal Ensemble (DE):
\begin{align}
&\rho_{DE} \equiv \sum_{\alpha} \left|c_{\alpha}\right|^2 \Ket{E_{\alpha}} \Bra{E_{\alpha}} \label{eq:DE}
\end{align}
where $c_{\alpha} \equiv \left\langle \psi_0 | E_{\alpha}\right\rangle$ and $\Ket{\psi_0}$ is the initial pure state of the isolated system. Roughly speaking, the two main assumptions made by Reimann are the following: first, that in any experimentally realistic condition the state of the system will occupy a huge number of energy eigenstates, even if the average energy is known up to a small experimental uncertainty; second, that the observable under study has a finite range of average values, due to the fact that we wish to measure it. For a clear and synthetic discussion on this topic we suggest \cite{Pure12} and we send the reader to the original references \cite{Rei1,Rei2}. We note that the first assumption does not contradict the small energy dispersion assumption. Indeed, as argued by Reimann, in a many-body quantum system, even if the energy is known up to a macroscopically small scale $\delta E$, there will be a huge number of eigenstates within the range $E_{\alpha} \in I_0$. This is formally equivalent to say that we will observe a high density of energy eigenstates. To conclude, given a HUB it is always possible to obtain a HUO which satisfies the finite-range assumption. We can therefore apply Reimann's theorem to HUOs.\\

It is important to note that Reimann's theorem explains equilibration around the DE but this does not necessarily entail thermalisation. The DE still retains information about the initial state while, on 
the contrary, thermalisation is defined (also) by the independence on the initial state. This is the point where our result is able to take a step forward and explain the emergence of thermal equilibrium in the HUOs. We can use Eq.(\ref{eq:EE}) to prove that all the HUOs exhibit complete independence from the initial conditions:

\begin{align}
& \mathrm{Tr} \left( \mathcal{O} \rho_{DE}\right) = \sum_{\alpha} |\psi_{\alpha}^{0}|^2 \mathcal{O}_{\alpha \alpha} = \mathrm{Tr} \left( \mathcal{O} \rho_{mc} \right) =\Mv{\mathcal{O}}_{mc} 
\end{align}

\subsubsection*{HUOs and ETH: off-diagonal matrix elements}

In order to prove that a HUO satisfies ETH we need to study also its off-diagonal matrix elements. By using Eq.(\ref{eq:Const}), we can investigate how the phases $\omega_{js}^{\alpha \beta} \equiv \left( \theta_{js,\beta} - \theta_{js,\alpha} \right)$ are distributed. This can be done numerically, exploiting  the available algorithms to generate MUBs\cite{MUB4}. The numerical investigation of the distribution of $\omega_{js}^{\alpha \beta}$ is reported in the supplementary material. Here we simply state the result: for a fixed value of the energy quantum numbers, the observed distributions of $\cos \omega_{js}^{\alpha \beta}$, $\sin \omega_{js}^{\alpha \beta}$ are well described by the assumption that $\omega_{js}^{\alpha \beta}$ are independent and randomly distributed in $\left[ -\pi,\pi \right]$, with a constant probability distribution.\\

There are different ways in which this result can be used. The general argument is the following and it reflects the spirit of ETH: in Eq.(\ref{eq:off}), the phases $\omega_{js}^{\alpha \beta}$ will have a randomising action on the eigenvalues $\lambda_j$  and this will make the value of the off-diagonal matrix elements severely smaller than the value of the diagonal ones:

\begin{align}
&\mathrm{Re} \mathcal{O}_{\alpha \beta}, \mathrm{Im} \mathcal{O}_{\alpha \beta} \ll \frac{1}{\mathcal{D}}\mathrm{Tr} \mathcal{O} =  \mathcal{O}_{\alpha \alpha}
\end{align}

The randomness of the coefficients involved in the evaluation of the off-diagonal matrix elements has been recently proposed as the basic mechanism to explain the $\frac{1}{\sqrt{\mathcal{D}}}$ scaling behaviour which was observed to occur in some models \cite{ETH18,ETH19,ETH20,ETH21,ETH22}. The maximisation of Shannon entropy is therefore giving us a recipe to find the observables for which this is true.\\

\subsubsection*{HUOs and ETH: two important examples}

In order to understand how this works in practice we need to say something specific about the eigenvalues. We highlight two important cases in which our result is helpful: an observable which is highly degenerate and an observable whose eigenvalues distribution is not correlated with the phases $\omega_{js}^{\alpha \beta}$.\\

{\it Highly degenerate observable} - Assuming that $j=1,\ldots, \mathcal{D}_1$ while $s=1,\ldots,\mathcal{D}_2$ with $\mathcal{D} \sim \mathcal{D}_2 \gg \mathcal{D}_1$, the sum in Eq. (\ref{eq:off}) splits into $\mathcal{D}_1$ of sums and each one of them is a sum of $\mathcal{D}_2 \gg 1$ identically distributed random variables. We can apply the central limit theorem to the real and imaginary part of Eq. (\ref{eq:off}) and obtain the following expression for the off-diagonal matrix elements 

\begin{align}
& \begin{array}{l}
\mathrm{Re} \mathcal{O}_{\alpha \beta}^{deg} \\
\mathrm{Im} \mathcal{O}_{\alpha \beta}^{deg}
  \end{array}  \approx  \sum_{j=1}^{\mathcal{D}_1} \lambda_j \mathcal{N}_{j} \left(0,\frac{1}{\sqrt{\mathcal{D}_2}}\right) \sim \frac{1}{\sqrt{\mathcal{D}}} \mathcal{R}_{\alpha \beta}
\end{align}

In which $\mathcal{N}(\bar{x},\sigma^2)$ indicates a gaussian probability distribution with mean $\bar{x}$ and variance $\sigma^2$ and $\mathcal{R}_{\alpha \beta}$ is a zero mean and unit variance random variable. This proves that a highly degenerate observable satisfies the ETH ansatz, Eq.(\ref{eq:ETH}). The diagonal matrix elements reproduces the microcanonical expectation values and the off-diagonal matrix elements are well described by a random variable with zero mean and  $\sim \frac{1}{\sqrt{\mathcal{D}}}$ variance. This is precisely what we expect from a local observable, since its eigenvalues have a huge number of degeneracies, which grows exponentially with the size of the system, and it is in full agreement with the randomness conjecture made in \cite{ETH21,ETH22}.\\

{\it Uncorrelated distribution} - If the eigenvalues distributions $\left\{ \lambda_j \right\}_{j,s}$ and the phases $\left\{e^{i \omega_{js}^{\alpha \beta}}\right\}_{j,s}$ are not correlated Eq.(\ref{eq:off}) becomes 

\begin{align}
&\mathcal{O}_{\alpha \beta}^{unc} \approx   \left(\sum_{j,s} \lambda_{j} \right) \left(\frac{1}{\mathcal{D}}\sum_{j,s} e^{i \omega_{js}^{\alpha \beta}}  \right) = \frac{\mathrm{Tr} \mathcal{O}}{\mathcal{D}} \,  \delta_{\alpha \beta}
\end{align}

Where we used the fact that the two sequences are uncorrelated to approximate the sum as the product of two sums and in the second identity we used the fact that $\left\{\Ket{j,s}\right\}$ is a complete basis. We conclude that the off-diagonal matrix elements of such an observable are much smaller than the diagonal ones and therefore we can neglect them when we compute its dynamical expectation value:

\begin{align}
&\Bra{\psi(t)} \mathcal{O}^{unc} \Ket{\psi(t)} = \sum_{\alpha,\beta} \psi_{\alpha}^{0} \bar{\psi}_{\beta}^{0} e^{- \frac{i}{\hbar} (E_{\alpha} - E_{\beta})t} \mathcal{O}_{\alpha \beta}^{unc}  \approx \sum_{\alpha} |\psi_{\alpha}^{0}|^2 \mathcal{O}_{\alpha \alpha}^{unc} = \Mv{\mathcal{O}^{unc}}_{mc}
\end{align}

 The assumption that the sequences $\left\{ \lambda_j \right\}_{j,s}$ and $\left\{e^{i \omega_{js}^{\alpha \beta}}\right\}_{j,s}$ are not correlated will not be always fulfilled, nevertheless this is what we intuitively expect from the following reasoning. We know that $\omega_{js}^{\alpha \beta}$ can be described by a random variable with a constant probability distribution, this suggests that there is no correlation between $\left\{\omega_{js}^{\alpha \beta}\right\}_{j,s}$ and its labeling $\left\{ j,s\right\}_{j,s}$. On the contrary, the eigenvalues $\lambda_j$ are highly correlated to the labels and therefore we do not expect them to be correlated to the $\left\{\omega_{js}^{\alpha \beta}\right\}_{j,s}$. \\

These results follows from the study of the equilibrium equations, under the assumption that the state is an Hamiltonian eigenstate $\Ket{\psi} = \Ket{E_{\alpha}}$ belonging to the energy shell $I_0$. It is straightforward to see that the same results hold when the state is the microcanonical state $\rho_{mc}(E_0,\delta E)$ involved in Eq.(\ref{eq:AverTherm}) and defined by the condition $\mathrm{Tr} \rho T \in I_0$.\\

In the next section we will exploit the proven results to investigate when local observables are expected to satisfy ETH, in a closed quantum system.

\section*{Discussion}

We proposed a new notion of thermal equilibrium for an observable $\mathcal{O}$: we say that $\mathcal{O}$ is a \emph{thermal observable} when its eigenvalues probability distribution maximises its Shannon entropy $H_{\mathcal{O}}$. Setting up a constrained optimisation problem we derived two equilibrium equations and studied their physical implications. Eq.(\ref{eq:EE1}) enforces the stability of the distribution with respect to the dynamics generated by the Hamiltonian while Eq.(\ref{eq:EE2}) fixes the functional form of the distribution. Integrating the second equation we obtained a linear relation between Shannon entropy and the mean value of the energy, which is a strong indication that $H_{\mathcal{O}}$ at equilibrium has some thermodynamic properties. The physical meaning of the two equilibrium equations provides evidence that the maximisation of Shannon entropy is able to address the emergence of thermal behaviour. We also studied the relation of the proposed notion of equilibrium with quantum statistical mechanics. We showed that it is possible to interpret the ordinary notion of thermal equilibrium through the maximisation of Shannon entropy. The request to maximise the lowest among all the possible Shannon entropies lead to Gibbs ensembles and therefore to the quantum statistical mechanics characterisation of thermal equilibrium.\\

Together, the physical meaning of the equilibrium equations and the proven relation between maximisation of Shannon entropy and Gibbs ensemble, provide strong evidence that the proposed notion of thermal observable is physically relevant for the purpose of investigating the concept of thermal equilibrium. This is our first result. \\

The result is the relevance of the proposed notion of equilibrium to address the emergence of thermal behaviour in a closed quantum system and especially its relation with the ETH. Using maximisation of $H_{\mathcal{O}}$ we were able to find a large class of observables which always thermalise and provide an algorithm to explicitly construct them. We call them Hamiltonian Unbiased Observables (HUOs). Using this result, we analysed the matrix elements of an HUO in the Hamiltonian eigenbasis to understand if they satisfy the ETH. The study of the diagonal matrix elements reveals that they always satisfy ETH, being constant in such basis. This first result on ETH has been used to prove thermalisation of the average value of HUOs, in connection with Reimann's theorem about equilibration of observables. Within the picture suggested by Reimann, the average value of observables can equilibrate around the diagonal ensemble, if certain conditions hold. Arguing that these conditions are satisfied by a HUO, we applied Reimann's theorem and obtained thermalisation of the average value of a HUO.\\

The study of the off-diagonal matrix elements of a HUO revealed that their value is exponentially suppressed, in the dimension of the Hilbert space. This complete the proof that HUOs always satisfy the ETH. The relevance of this result for the pure-states statical mechanics is related to the two main objections usually raised against ETH: the lack of predictive power for what concern both which observables satisfy ETH and how long it should take them to reach thermal equilibrium. The proposed notion of thermal equilibrium is therefore revealing its predictive power since it gives us a way of finding observables which always satisfy ETH, in a closed quantum system.\\

We would like to conclude by putting this set of results about ETH in a more general perspective. ETH is one of the main paradigms to justify the applicability of statistical mechanics to closed many-body quantum systems. However, it is just a working hypothesis, it is not derived from a conceptually clear theoretical framework. For such a reason, one of the major open issues is its lack of predictability. Despite that, there has been a huge effort to investigate whether the ETH can be invoked to explain thermalisation in concrete Hamiltonian models \cite{ETH14,ETH15,ETH16,ETH17,ETH18,ETH19,ETH20,ETH21,ETH22,ETH23,ETH24,ETH25,ETH26,ETH28} and it's use it is nowadays ubiquitous. Therefore, we think it is very important to put the ETH under a conceptually clear framework. In this sense, the relevance of our work resides in the fact that we obtain the ETH ansatz as a prediction, by using a natural starting point, the maximisation of Shannon entropy. Furthermore, using the proposed notion of thermal equilibrium is already giving concrete benefits. First, we now have a way of computing observables which satisfy the ETH and this prediction can be tested both numerically and experimentally; second, by studying the conditions under which local observables satisfy our equilibrium equations we are able to give two conditions which are necessary to guarantee the validity of ETH for all local observables. We believe that our investigation strongly suggest that maximisation of $H_{\mathcal{O}}$ is able to grasp the main intuition behind ``thermalisation according to ETH'' and it can be the physical principle behind the appearance of ETH.\\ 

Further investigation in this direction is certainly needed, but we would like to conclude by suggesting a way in which this new tool can be used to address the long-standing issue of the thermalisation times. From our investigation one can infer that Shannon entropy is a good figure of merit to study the dynamical onset of thermalisation in a closed quantum system. Within this picture, the time-scale at which thermalisation should occur for $\mathcal{O}$ is therefore given by the time-scale at which $H_{\mathcal{O}}$ reaches its maximum value. A prediction about the time-scale at which  $H_{\mathcal{O}}$ is maximised will translate straightforwardly in a prediction about the thermalisation time.\\

\section*{Methods}

Here we present the full derivation of the equilibrium equations, using the Lagrange multipliers' technique. Our purpose is to find the distribution which maximises Shannon entropy of an observable $\mathcal{O}$. The space on which such an optimisation is formulated is the space of the finite-dimensional density matrices, which is the space of positive semi-definite and self-adjoint matrices. The maximisation is constrained by two equations: the first one accounts for the normalisation of the state while the second one accounts for a fixed value of the average energy. \\

If one wants to be absolutely rigorous, there is another constraint which needs to be imposed, which is the non-negativity of the density matrix. This is commonly indicated in the following way: $\rho \geq 0$. In other words, using the above-mentioned constraints there is no guarantee that, if a solution exists, it is a positive density matrix. This is not an actual problem, from the physical point of view, because one simply disregards a solution if it does not give a non-negative matrix. In such a case, the mathematical problem might have a solution, but not the associated physical question, which is the one we are really interested in. We would like to suggest an alternative route in such direction. If one wants to be completely rigorous, the Karush-Kuhn-Tucker (KKT) technique can be used to implement the non-negativity constraint on the density matrix.  However, we do not think we would gain any physical insight from the use of such technique and therefore we are not going to explore such a route.\\

We present here the derivation, in the general case of a mixed state $\rho = \sum_{n} q_n \Ket{\psi_n}\Bra{\psi_n}$ and of a degenerate observable $\mathcal{O} = \sum_{j,s} \lambda_j \Ket{j,s} \Bra{j,s}$, in which $\left\{ \Ket{j,s} \right\}$ is a complete basis of the Hilbert space. Here are the two constraints:

\begin{subequations}
\begin{align}
&\mathcal{C}_N \equiv \mathrm{Tr}\left( \rho\right) - 1 =\sum_{n;j,s} q_n |D_{js}^{(n)}|^2 -1 \\
&\mathcal{C}_E \equiv \mathrm{Tr} \left(\rho T \right) - E_0 = \sum_{n;j,s;j',s'} q_n \overline{D}_{js}^{(n)} T_{js,j' s'} D_{j' s'}^{(n)} - E_0
\end{align}
\end{subequations}

in which $D_{js}^{(n)}\equiv \BraKet{j,s}{\psi_n}$. Moreover $T_{js;j' s'} = \Bra{j,s}\hat{T}\Ket{j',s'}$. Exploiting Lagrange's multipliers technique one defines an auxiliary function $\Lambda_{\mathcal{O}}$, specific for the $\mathcal{O}$ observable, that can be freely optimised

\begin{align}
&\Lambda_{\mathcal{O}} [\hat{\rho},\lambda_E,\lambda_N] \equiv H_{\mathcal{O}}[\hat{\rho}] + \lambda_N \mathcal{C}_N + \lambda_E \mathcal{C}_E
\end{align}

The derivatives with respect to the Lagrange's multipliers $\lambda_E$ and $\lambda_N$ enforce the validity of the constraints 

\begin{subequations}
\begin{align}
&\frac{\delta \Lambda_{\mathcal{O}}}{\delta \lambda_N} = 0  \quad \Rightarrow \quad \mathcal{C}_N = 0\\
&\frac{\delta \Lambda_{\mathcal{O}}}{\delta \lambda_E} = 0  \quad \Rightarrow \quad \mathcal{C}_E = 0
\end{align}
\end{subequations}

while the derivatives with respect to the field variables and with respect to the statistical coefficients $q_n$ gives three equations, of which only two are independent:

\begin{align}
&\frac{\delta \Lambda_{\mathcal{O}}}{\delta D_{js}^{(n)}} = 0 && \frac{\delta \Lambda_\mathcal{O}}{\delta \bar{D}_{js}^{(n)}} = 0  
\end{align}

where

\begin{subequations}
\begin{align}
&\frac{\delta \Lambda_{\mathcal{O}}}{\delta D_{jn}} = - q_n \bar{D}_{js}^{(n)} \left[ \log \left( \sum_{m,p} q_m |D_{jp}^{(m)}|^2 \right) + (1-\lambda_N) \right] + \lambda_E q_n \sum_{i,p} T_{js,ip} \bar{D}_{ip}^{(n)}\\
&\frac{\delta \Lambda_\mathcal{O}}{\delta \bar{D}_{jn}} = - q_n D_{js}^{(n)} \left[ \log \left( \sum_{m,p} q_m |D_{jp}^{(m)}|^2 \right) + (1-\lambda_N) \right] + \lambda_E q_n \sum_{i,p} D_{ip}^{(n)} T_{ip,js} 
\end{align}
\end{subequations}

Instead of using these equations, we use the two following independent linear combinations

\begin{subequations}
\begin{align}
&D_{js}^{(n)} \frac{\delta \Lambda_{\mathcal{O}}}{\delta D_{js}^{(n)}} - \bar{D}_{js}^{(n)} \frac{\delta \Lambda_{\mathcal{O}}}{\delta \bar{D}_{js}^{(n)}}= 0  \\
& \frac{1}{2} \left(D_{js}^{(n)} \frac{\delta \Lambda_{\mathcal{O}}}{\delta D_{js}^{(n)}} + \bar{D}_{js}^{(n)} \frac{\delta \Lambda_{\mathcal{O}}}{\delta \bar{D}_{js}^{(n)}}= 0    \right)= 0 
\end{align}
\end{subequations}

which give the two equilibrium equations that we used in the main text

\begin{subequations}
\begin{align}
&\overline{\mathcal{E}}_n(j,s) = \mathcal{E}_n(j,s) \\
&- |D_{js}^{(n)}|^2 \log \left(\sum_{m,q} q_m  |D_{jq}^{(m)}|^2\right)  =   \left( 1-\lambda_N \right)|D_{js}^{(n)}|^2    -  \lambda_E \mathcal{E}_n(j,s) 
\end{align}
\end{subequations}

\section*{Acknowledgements}

F.A. would like to thank the ``Angelo Della Riccia'' foundation and the St. Catherine's College, University of Oxford for their support to this research. V.V. acknowledges funding from the John Templeton Foundation, the National Research Foundation (Singapore), the Ministry of Education (Singapore), the Engineering and Physical Sciences Research Council (UK), the Leverhulme Trust, the Oxford Martin School, and Wolfson College, University of Oxford. This research is also supported by the National Research Foundation, Prime Ministers Office, Singapore under its Competitive Research Programme (CRP Award No. NRF- CRP14-2014-02) and administered by Centre for Quantum Technologies, National University of Singapore.  

\section*{Author contributions statement}

F.A. proposed the idea. V.V. contributed to the development. Both authors reviewed the manuscript. F.A. would like to thank Dr. Davide Girolami for several discussions on this topic.

\section*{Additional information}

Competing financial interest: \textbf{The authors declare no competing financial interests}

\end{document}